\documentclass[preprint,preprintnumbers,a4paper,nofootinbib]{revtex4-1}
\usepackage{amsmath,amssymb}
\usepackage{graphicx}
\usepackage[utf8]{inputenc}
\usepackage{color}

\newcommand{\Dcut}{D_{\mathrm{cut}}}
\newcommand{\Tc}{T_{\mathrm{c}}}

\begin{document}

\begin{flushright}%
{\small UTHEP-729, UTCCS-P-119, HUPD-1809}
\end{flushright}%

\begin{center}
{\Large{\textbf{ Irregular parameter dependence of numerical results in 
tensor renormalization group analysis}}}
\\
\medskip
\vspace{1cm}
\textbf{
Daisuke Kadoh$^{\,a,b,}$\footnote{kadoh@keio.jp}, 
Yoshinobu Kuramashi$^{c,}$\footnote{kuramasi@het.ph.tsukuba.ac.jp}, 
Ryoichiro Ueno$^{d,}$\footnote{ryoichiro-ueno@hiroshima-u.ac.jp}
}
\bigskip

$^a$ {\small Department of Physics, Faculty of Science, Chulalongkorn University, Bangkok 10330, Thailand}

$^b$ {\small Research and Educational Center for Natural Sciences, Keio University, \\ Yokohama 223-8521, Japan}

$^c$  {\small Center for Computational Sciences, University of Tsukuba, Tsukuba 305-8577, Japan}

$^d$ {\small  Graduate School of Science, Hiroshima University, Higashi-Hiroshima 739-8526, Japan}

\end{center}

\begin{abstract}
We study the parameter dependence of numerical results obtained by the tensor renormalization group. 
We often observe an irregular behavior as the parameters are varied with the method, 
which makes it difficult to perform the numerical derivatives in terms of the parameter.
With the use of two-dimensional Ising model we explicitly show that the sharp cutoff used in the truncated singular value decomposition causes this unwanted behavior when the level crossing happens between singular values below and above the truncation order as the parameters are varied.
We also test a smooth cutoff, instead of the sharp one, as a truncation scheme and discuss its effects. 
\end{abstract}

\maketitle

\section{Introduction}

Tensor renormalization group (TRG) is a promising approach that can
solve the sign problem inherent in the Monte-Carlo simulations. 
Since it was proposed in two-dimensional Ising model \cite{Levin:2006jai},  
many studies have been carried out for various models of lattice field theories \cite{Shimizu:2012wfa,Shimizu:2014uva,Unmuth-Yockey:2014iga,Shimizu:2014fsa,Takeda:2014vwa,Kawauchi:2016xng,Meurice:2016mkb,Sakai:2017jwp,Yoshimura:2017jpk,Shimizu:2017onf,Kuramashi:2018mmi,Kadoh:2018hqq}.
In TRG, the truncated singular value decomposition (SVD) 
is used to define a coarse-grained tensor, 
which is given in a manner of sharp cut-off such that 
the $\Dcut$ largest singular values and corresponding singular vectors are kept and the others are thrown away.
Although the cutoff yields possible systematic errors, 
it is expected that the result should converge to the correct value as  $\Dcut$ increases.

The results of TRG, however, do not smoothly depend on the parameters in the theory.  
They often show irregular behavior at some parameters off the critical point. 
This behavior is controlled by $\Dcut$, 
but for small  $\Dcut$ it is difficult to obtain the smooth parameter dependence of the results, 
to which we may apply
the numerical derivative with respect to the parameter.
We can of course obtain a satisfactory result for a simple model such as  
two-dimensional Ising model taking a sufficiently large   $\Dcut$
to avoid such misbehavior. 
However, it would be difficult to increase  $\Dcut$ for general lattice theories 
with multi-dimensional fields so that it should be important to understand 
and avoid the irregular behavior of the results.

In this paper we investigate the origin of the irregular parameter dependence 
shown in the TRG results. 
We present some numerical evidence that it is caused 
by the level crossing between singular values within and beyond the sharp truncation 
as the parameters are varied.
In this sense the irregular behavior is inevitable for the TRG method with the sharp cutoff. 
In order to obtain a hint of improving the behavior, we also test other cutoff schemes 
such as a smooth cutoff.

The rest of this paper is organized as follows. 
In Sec.~\ref{sec:trg} we review the standard TRG method with the sharp cutoff in two-dimensional Ising model 
with sample numerical results.
The mechanism of irregular behavior is explained in detail with some numerical evidence in Sec.~\ref{sec:results}. 
We also test other cutoff schemes. 
Our conclusions are summarized in Sec.~\ref{sec:summary}.

\section{TRG in two dimensional Ising model}
\label{sec:trg}

We briefly review the TRG method in two-dimensional Ising model 
presenting a couple of numerical results for later convenience. 

\subsection{Numerical procedures in TRG}
We consider a two-dimensional square lattice whose sites are labeled by $n=(n_1,n_2)$ for $n_1,n_2 \in \mathbb{Z}$. 
The spin variable  $\sigma_n$ assigned on the site $n$ takes the discrete values $\sigma_n \in \{1,-1\}$. 
Two dimensional Ising model is then defined by the Hamiltonian
\begin{align}
	{\cal H}=-J \sum_{\langle i,j \rangle}\sigma_{i}\sigma_{j},
\end{align}
where $\langle i,j \rangle $ denotes possible pairs of the nearest neighbor sites and $J$ is the coupling constant.

The partition function $Z={\rm Tr}\, {\rm e}^{-\beta {\cal H}}$ with the inverse temperature $\beta=1/T$ can be expressed as a tensor network form:
\begin{eqnarray}
Z=\sum_{i,j,k,l,\dots} T_{ijkl} \ldots
	\label{tensor_representation}
\end{eqnarray}
where 
\begin{eqnarray}
\label{initial_tensor}
T_{ijkl} = e^{\beta J(ij + j k + kl + li)}
\end{eqnarray}
for $i,j,k,l=-1,1$.

Let us denote the bond dimension of $T_{ijkl}$ as $N$ for the sake of argument. 
Note that the initial tensor of Eq.~(\ref{initial_tensor}) is defined with $N=2$.   
We apply the truncated SVD to $T_{ijkl}$:
\begin{align}
	T_{ijkl} &  \approx \sum_{m=1}^{\Dcut }U_{(ij)m}\lambda_{m}V^{\dagger}_{m(kl)}
	\label{SVD1}
	\\
	T_{ijkl}   & \approx \sum_{m=1}^{\Dcut }U^{\prime}_{(li)m}\lambda^{\prime}_{m}{V^{\prime}}^{\dagger}_{m(jk)},
	\label{SVD2}
\end{align}
where $T_{ijkl}$ is treated as a matrix with the column $(ij)$ and row $(kl)$ in Eq.~(\ref{SVD1}) and  
a matrix with the column $(li)$ and row $(jk)$ in Eq.~(\ref{SVD2}).
The above expressions assume the case of $N^2 >  \Dcut $, 
while $\Dcut $ in Eqs.~(\ref{SVD1}) and (\ref{SVD2})  is replaced by $N^2$ for $N^2 \le \Dcut $ without any truncation. We apply the decomposition of Eq.~(\ref{SVD1}) to the tensors at even sites defined by mod($n_1+n_2$,2)=0 and that of Eq.~(\ref{SVD2}) to ones at odd sites with mod($n_1+n_2$,2)=1.
Here $U,V,U^{\prime},V^{\prime}$ are unitary matrices and $\lambda_m$ and $\lambda'_m$ are singular values 
that are sorted in descending order.

We immediately find that the expression of Eq.~(\ref{tensor_representation}) is approximated as 
\begin{eqnarray}
Z \approx \sum_{i,j,k,l,\dots} T^{\rm new}_{ijkl}\ldots,
	\label{new_tensor_representation}
\end{eqnarray}
where 
\begin{eqnarray}
\label{new_tensor}
T^{\rm new}_{ijkl}= \sqrt{\lambda_i \lambda^\prime_j \lambda_k \lambda^\prime_l}
\sum_{a,b,c,d=1}^{\Dcut}  U_{(ab)i}  U^{\prime}_{(bc)j} V^{\dagger}_{(cd)k}  V^\dag_{(da)l}.
\end{eqnarray}
Note that the number of tensors decreases because an old tensor is decomposed into two unitary matrices 
$U$ and $V$ (or $U'$ and $V'$) and then four unitary matrices are assembled into a new tensor.

After repeating the above procedures, the tensor network is finally reduced to a single tensor.
Taking the appropriate trace for its indices we obtain the approximate value of $Z$ with $\Dcut$. 
The numerical cost of this algorithm is  $O(\Dcut^{6})$ 
which comes from the computations of
Eqs.~(\ref{SVD1}) and (\ref{SVD2}).

\subsection{Numerical examples with TRG analysis}

TRG is a powerful tool to study two-dimensional lattice models. 
Although the exact value is obtained in the $\Dcut \rightarrow \infty$ limit,
we can reach a sufficient level of accuracy with moderate number of $\Dcut$ in practical computations. 
We present a couple of representative results in the TRG analysis for 
two-dimensional Ising model on $V=2^{16}\times 2^{16}$ lattice
as a preparation of our study explained  in the following section.

The numerical value of the partition function $Z$ is obtained by 
repeating the renormalization step of TRG with a given value of $\Dcut$. 
Then the Helmholtz free energy $F$ is also obtained with the use of $F=-T\log(Z)$.
The critical temperature $T_c$ is determined from the peak position of the specific heat $C_V$  
obtained by the numerical derivative of $Z$ with respect to $\beta$ as 
$C_V= -\beta^2\frac{\partial^2}{\partial \beta^2} {\rm log} Z$.

\begin{figure}[t]
	\begin{center}
		\includegraphics[clip,scale=0.8]{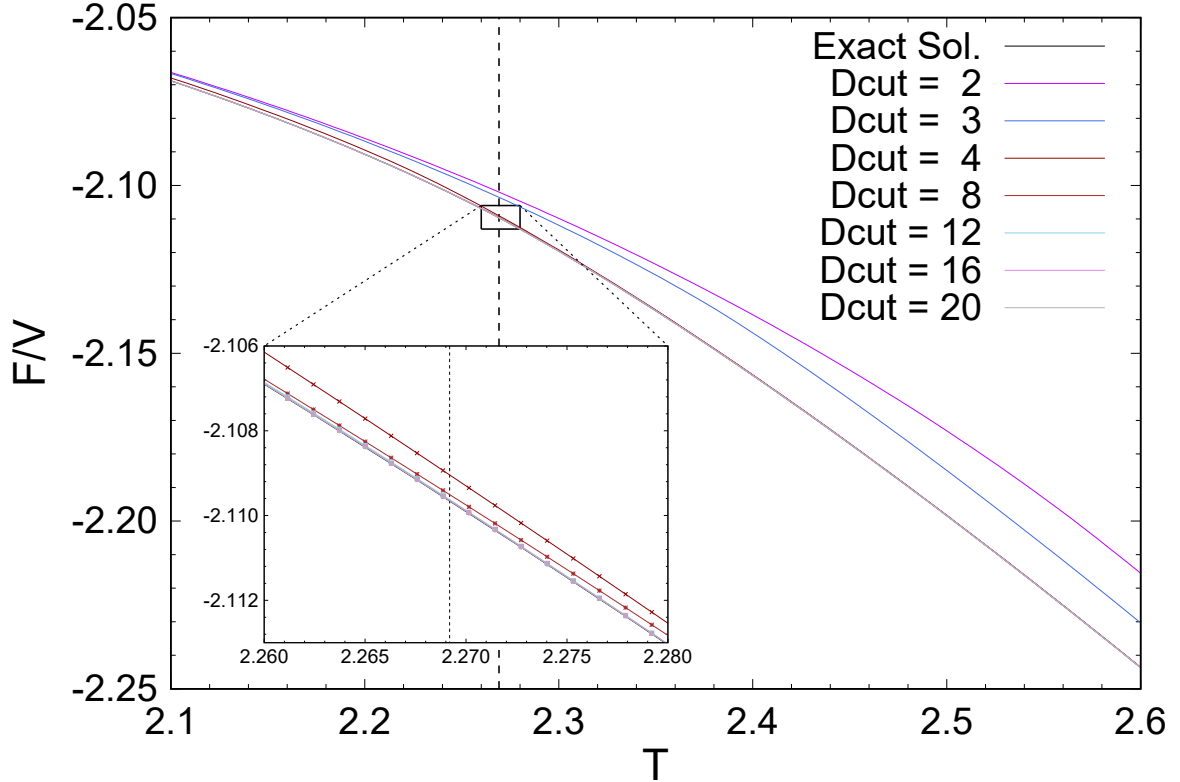}
		\caption{
		$T$ dependence of free energy density evaluated on $V=2^{16}\times 2^{16}$ lattice.
		}
		\label{free_energy}
	\end{center}
\end{figure}

Figure~\ref{free_energy} shows the temperature dependence of the free energy density.
The black curve denotes the exact solution given in Ref.~\cite{PhysRev.65.117,PhysRev.76.1232}, and the black dotted line 
denotes 
the critical temperature. 
As clearly seen in the figure, the TRG results approach the exact solution 
as the value of $\Dcut $ increases.  
The results with $\Dcut \geq 4$ reproduce the exact one 
within the error of the order of $10^{-5}$.

\begin{figure}[t]
	\begin{center}
		\includegraphics[clip,scale=0.8]{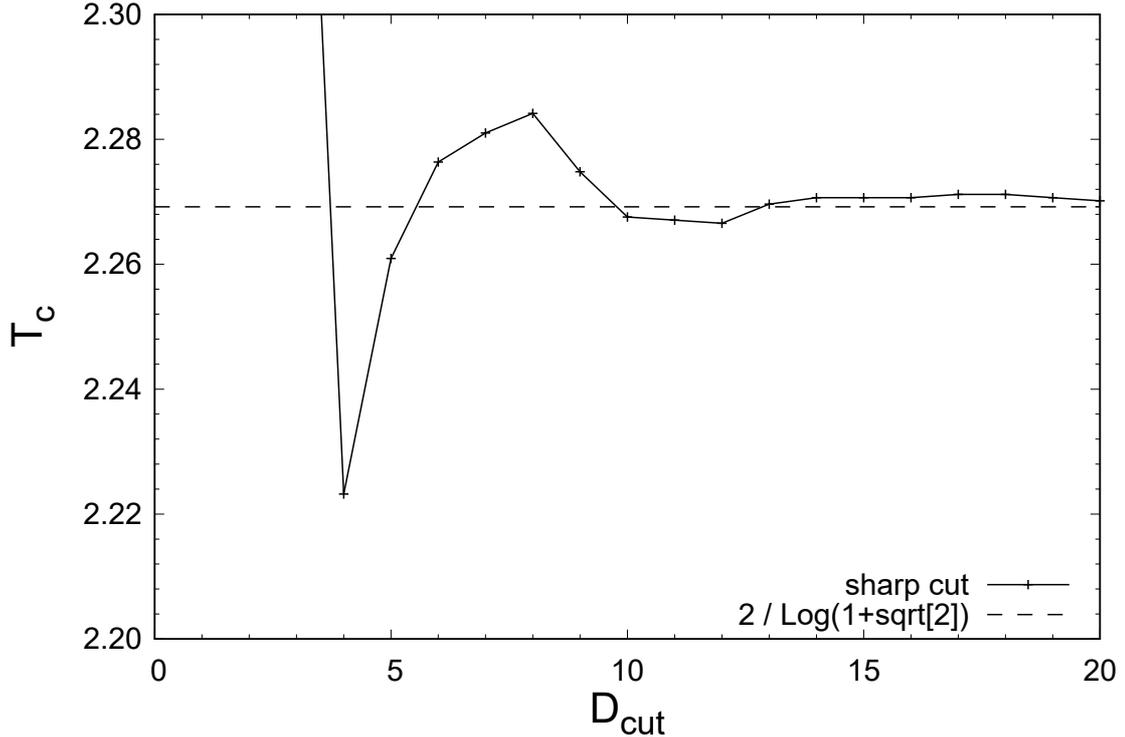}
		\caption{
         Critical temperature evaluated on $V=2^{10} \times 2^{10}$ lattice. The computational 
         results are presented as black points, and a dotted line denotes the exact value of the critical temperature 
         $\Tc^{exact} = 2/[\log(1+\sqrt{2})]=2.2691853\cdots$.}
		\label{Dcut_dependence_of_Tc}
	\end{center}
\end{figure}

Figure~\ref{Dcut_dependence_of_Tc} 
shows the $\Dcut$-dependence of the critical temperature.  
The numerical results fluctuate around the exact solution $T_c^{\rm exact}=2/[\log(1+\sqrt{2})]=2.2691853\cdots$.
 It is clearly observed that taking the enlarged value of $\Dcut$ 
makes the results approach the exact one.

%
%
%
\section{Irregular parameter dependence of TRG and new scheme with smooth cut}
\label{sec:results}

\subsection{Origin of irregular behavior}

The numerical results of TRG often show the irregular behavior as the parameters are varied.
Here we consider the reason why the numerical results do not smoothly depend  on the parameters. 
For simplicity, the numerical computations are performed on $V=(16)^2$ in this section.

\begin{figure}[t]
	\begin{center}
		\includegraphics[clip,scale=0.8]{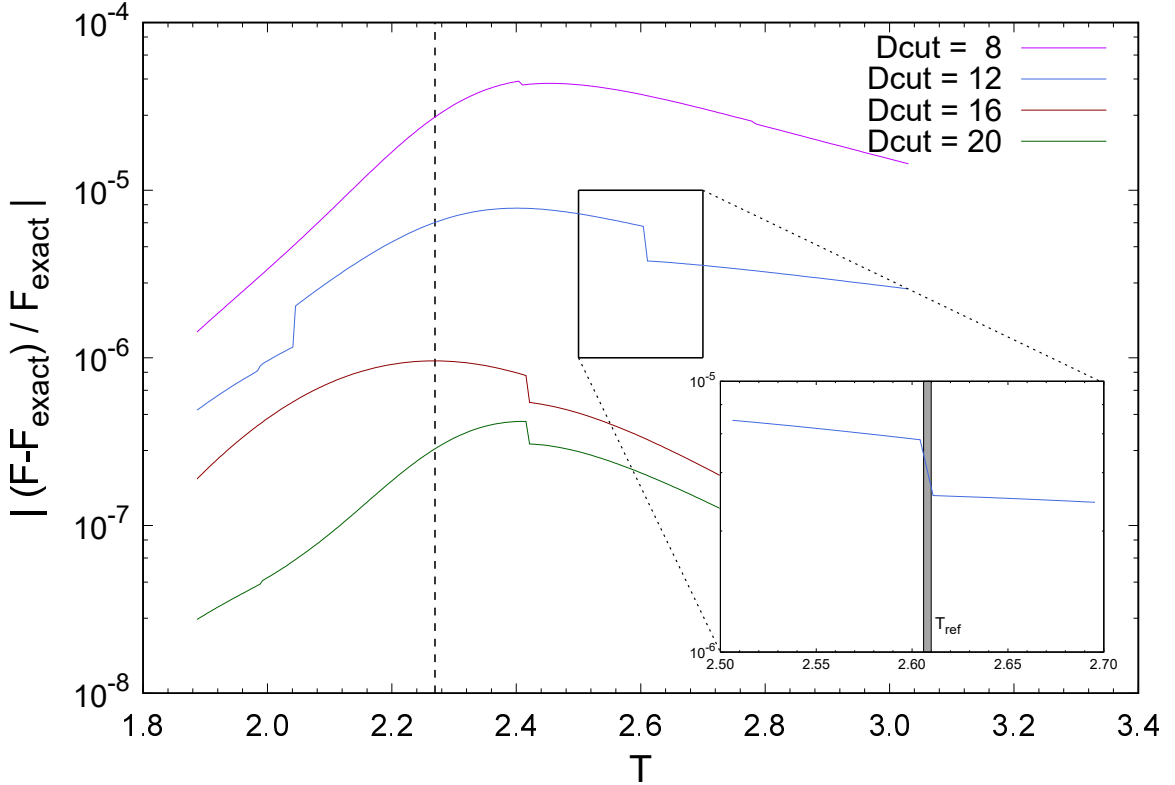}
		\caption{
		$T$-dependence of the relative residual of the Helmholtz free energy, 
         which is evaluated for $\Dcut=8, 12, 16,20$ on $V=(16)^2$ lattice.
		The gray band in an enlarged figure at the bottom right denotes the region
		where the irregular behavior is caused.
		}
		\label{crossover}
	\end{center}
\end{figure}

Figure~\ref{crossover} shows the relative residue of the free energy, 
which is given by a relative difference between the results of TRG 
and the exact solution.
The irregular behavior is observed as the abrupt jump of the results 
at several temperatures off the critical point denoted by the black dotted line.
For instance, as magnified in the small figure, the result with $\Dcut=12$ 
shows an irregular jump at $T_{\rm ref} \approx 2.6075$.

To understand the origin of this behavior,
we write the R.H.S. of Eq.~(\ref{SVD1}) as a form,  
\begin{align}
	T^{(1)}_{IJ} = \sum_{m=1}^{\Dcut}\lambda_{m} {u}^{(m)}_I {v}^{(m)}_{J},
	\label{approximated_tensor1}
\end{align}
where $\vec{u}^{(m)}$ ($\vec{v}^{(m)}$) is the left (right) singular vector corresponding 
to the $m$-th singular value $\lambda_{m}$. 
$T^{(1)}_{IJ}$ is an approximation of the tensor $T_{ijkl}$ for $I=(i,j)$ and  $J=(i,j)$.
We assume that a set of singular values and corresponding singular vectors 
$(\lambda, \vec{u}(\lambda), \vec{v}(\lambda))$ 
smoothly change under a local variation of parameters. 

We now consider a case in which the level crossing takes place 
such that $K$-th and $(K+1)$-th singular values are interchanged at some value of the parameter. 
The crossover is not important in the case of $K \neq \Dcut$ because 
both of $K$-th and $(K+1)$-th singular values are included (or not included)
in Eq.~(\ref{approximated_tensor1}). 
In the case of $K=\Dcut$, however, the crossover could make Eq.~(\ref{approximated_tensor1}) 
change drastically: the $\Dcut$-th singular vector before the crossover becomes the $(\Dcut+1)$-th one 
after the crossover and vice versa, while the $\Dcut$-th singular value changes continuously.

\begin{figure}[t]
	\begin{center}
		\includegraphics[clip,scale=0.8]{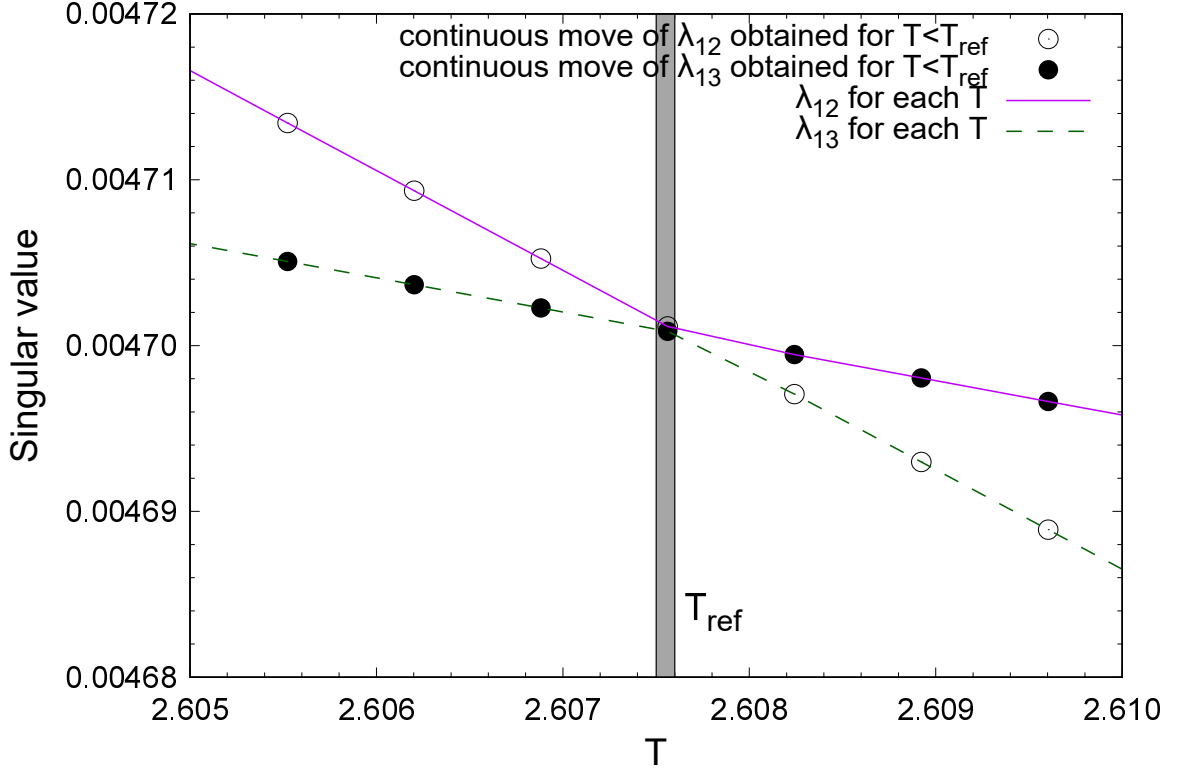}
		\caption{
		An example of the crossover of the $\Dcut$-th and $(\Dcut+1)$-th singular values, where $\Dcut=12$, as a function of $T$. 
		See text for lines and points.
		}
		\label{crossover2}
	\end{center}
\end{figure}

In Fig.~\ref{crossover2}
we trace the continuous move of the $\Dcut$-th and $(\Dcut+1)$-th singular values 
around $T_{\rm ref}$ (gray band), where  the $\Dcut$-th singular value (open circle) 
and  the $(\Dcut+1)$-th singular value (solid circle) 
are interchanged. 
Those singular values are obtained  in the course-grained tensor after six renormalization steps  
with $\Dcut=12$.  
We should note that the $(\Dcut+1)$-th singular value becomes 
 the $\Dcut$-th singular value after the level crossing.
The purple line is the $\Dcut$-th singular value included in Eq.~(\ref{approximated_tensor1}), 
and the green dotted line is the $(\Dcut+1)$-th one. 
This behavior suggests that the discontinuity of the result of the free energy does not 
come from the singular values in Eq.~(\ref{approximated_tensor1})
but from the discontinuous change of the singular vectors.

Let us consider the following modification for the approximation of tensor at the final step of SVD: 
\begin{eqnarray}
	T^{(2)}_{IJ} =
	\left\{ 
\begin{array}{ll}
T^{(1)}_{IJ}  & \quad {\rm for} \ T < T_{\rm ref}, \\
 \displaystyle \sum_{m=1}^{\Dcut-1}\lambda_{m} {u}^{(m)}_I {v}^{(m)}_{J} + \lambda_{\Dcut+1} {u}^{(\Dcut+1)}_I {v}^{(\Dcut+1)}_{J}   & \quad {\rm for} \  T \ge T_{\rm ref}. 
\end{array}
\right.
	\label{approximated_tensor2}
\end{eqnarray}
The meaning of this approximation is obvious from the definition. 
$T^{(2)}$ coincides with  $T^{(1)}$ before the level crossing. After the level crossing, however,
$T^{(2)}$ continues to keep the same sets $(\lambda, \vec{u}(\lambda), \vec{v}(\lambda))$ unlike $T^{(1)}$.  
If the irregular behavior is caused 
by the change of the associated singular vectors, 
it is expected that the jump at $T_{\rm ref}$ should vanish with the use of $T^{(2)}$.

\begin{figure}[t]
	\begin{center}
		\includegraphics[clip,scale=0.8]{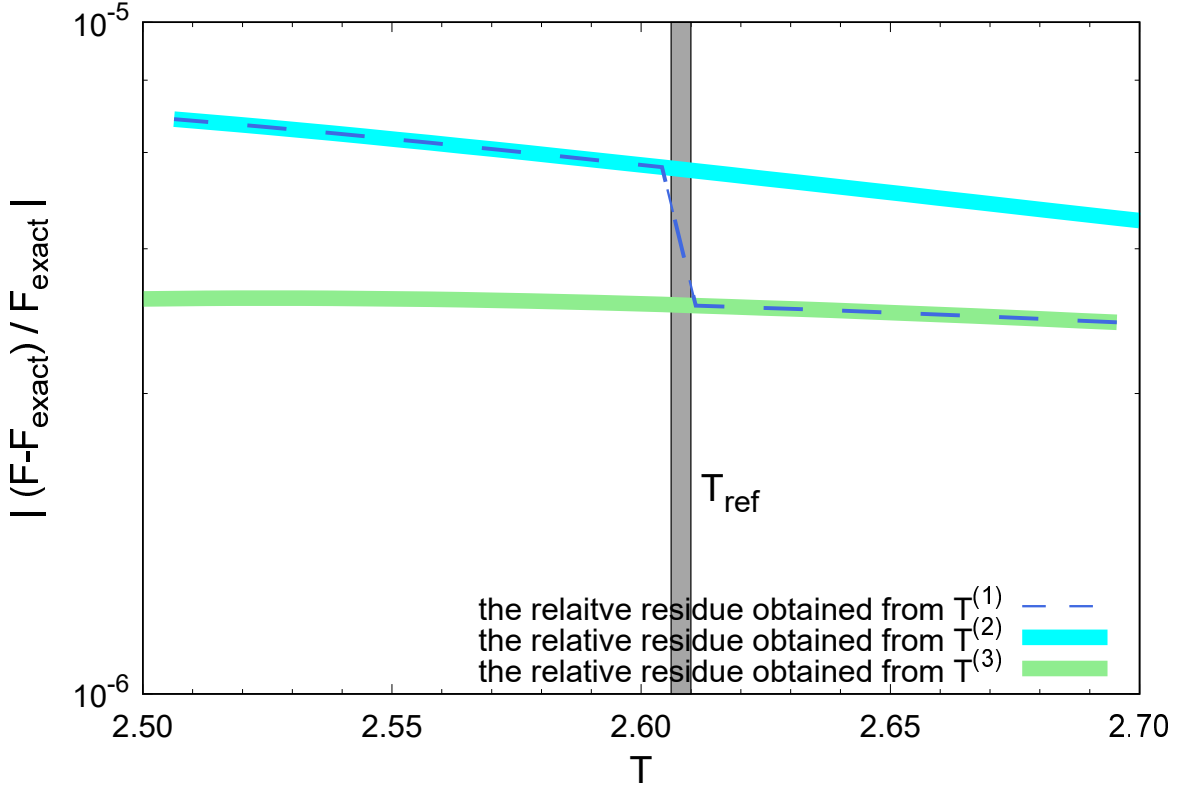}
		\caption{
		$T$-dependence of the relative residue of the free energy evaluated 
        with $\Dcut=12$ on $V=(16)^2$ lattice.
       The blue broken line, the blue and green curves are the relative residue 
        of the free energy obtained from $T^{(1)}$, $T^{(2)}$ and $T^{(3)}$, respectively.
		$T^{(1)}$ is defined by Eq.~(\ref{approximated_tensor1}), and $T^{(1)}$ is defined by Eq.~(\ref{approximated_tensor2}).
		$T^{(3)}$ is the case that the $(\Dcut+1)$-th set is used instead of $\Dcut$-th set
		for $T<T_\mathrm{ref}$ and $T^{(1)}$ is used for $T\ge T_{\mathrm{ref}}$.
		}
		\label{crossover3}
	\end{center}
\end{figure}

Figure~\ref{crossover3} shows the residues obtained from $T^{(2)}$, which are drawn by the blue curve. 
They smoothly depend on temperature and the jump at $T_{\rm ref}$ has gone.  
It is also instructive to check the smooth behavior of the green curve, which represent 
the results in the case that 
the $(\Dcut+1)$-th set is used instead of $\Dcut$-th set for $T < T_{\rm ref}$ (and $T^{(1)}$ is used for $T \ge T_{\rm ref}$).
We thus conclude that the irregular behavior is caused 
by the level crossing of the $\Dcut$-th and $(\Dcut+1)$-th singular values.
More specifically speaking, the replacement of the $\Dcut$-th singular vector at the crossover point 
yields the jump of the results.

\subsection{Test of smooth cutoff schemes}

The irregular parameter dependence of the results obtained by the TRG method is caused 
by the level crossing of the singular values across the truncation order as explained in the previous section. 
The standard TRG employs the sharp cutoff  
such that the  $\Dcut$ largest singular values and the associated vectors 
are included in the renormalization steps and the others are thrown away. 
In this section, we test other cutoff scheme such as a smooth cutoff 
to tame the misbehavior.

In order to define another truncation scheme,
we introduce a weight function $w_m$ to approximate the tensor $T_{ijkl}$:
\begin{align}
	T_{ijkl}\simeq\sum_{m=1}^{\Dcut}  w_{m} U_{(ij)m}\lambda_{m}V^{\dagger}_{m(kl)},
	\label{scut}
\end{align}
where $U$ and $V$ are unitary matrices and $\lambda_m$ are singular values sorted in descending order.
Note that Eq.~(\ref{SVD1}) is given by the choice of the weight function,
$w_m=1$ for $m\le \Dcut$. It can be expected that
the crossover effect depends on $w_m$ and may become weaker if we employ a smoother cutoff function
for $w_m$.
Note that  the introduction of $w_m$ itself does not demand extra computational cost. 

\begin{figure}[t]
	\begin{center}
		\includegraphics[clip,scale=0.8]{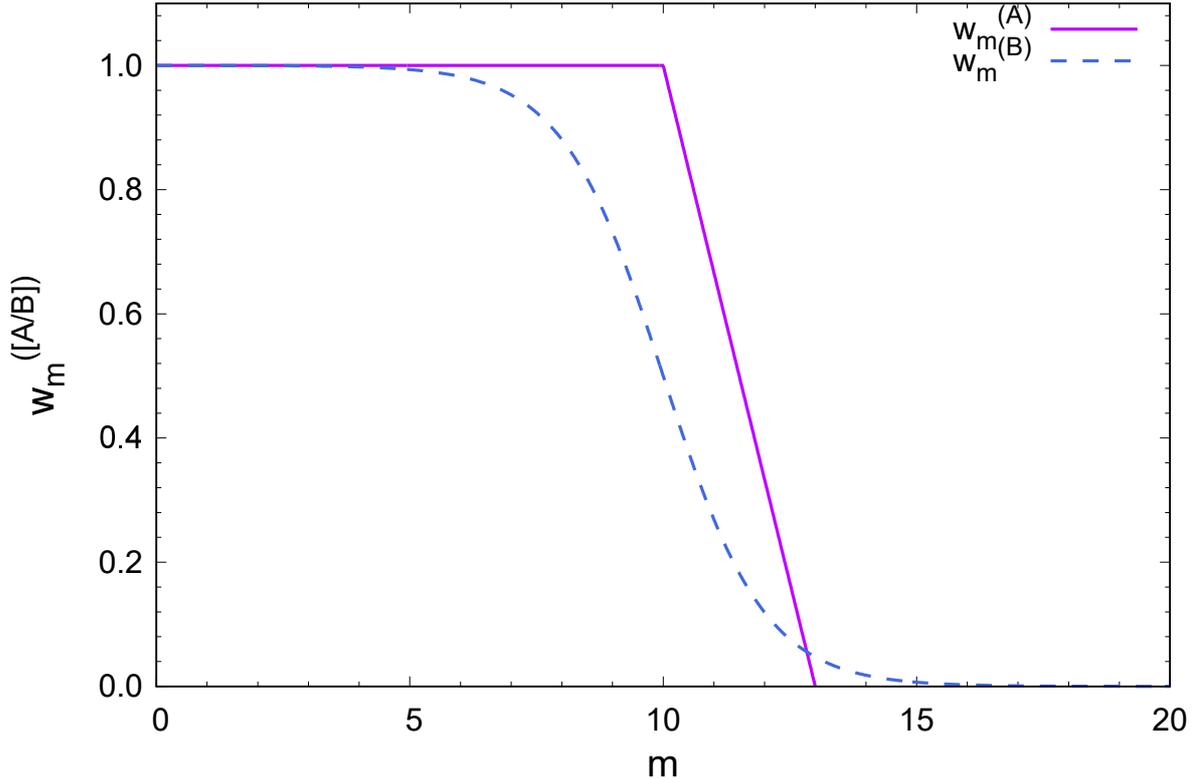}
		\caption{Weight factors $w_{m}^{(\mathrm{A})}$ and $w_{m}^{(\mathrm{B})}$.}
		\label{weight_factor}
	\end{center}
\end{figure}

As possible choices of cutoff schemes  we consider two types of weight functions:
(A) "a slanting-cut" given by
\begin{eqnarray}
w_m^{\rm (A)}=
	\begin{cases}
		1&(1\leq m\leq\Dcut - \Delta)
		\\
		\frac{\Dcut - m}{\Delta} & (\Dcut - \Delta <m\leq\Dcut), 
	\end{cases}
	\label{eq:smoothcut_a}
\end{eqnarray}
and  (B) "a FDF-cut" inspired by by the Fermi distribution function
\begin{align}
	w^{\rm (B)}_{m} = \frac{1}{\mathrm{e}^{(m-\Dcut)/\sigma}+1}.
	\label{eq:smoothcut_b}
\end{align}
Figure~\ref{weight_factor} shows examples of $w_{m}^{\rm (A)}$ and $w_{m}^{\rm (B)}$. 
$\Delta$  in $w^{\rm (A)}$ and $\sigma$ in $w^{\rm (B)}$ are the tunable parameters which basically give the smeared size of cutoff. 
In this paper we employ $\Delta=3$ and $\sigma=1$.

\begin{figure}[t]
	\begin{center}
		\includegraphics[clip,scale=0.8]{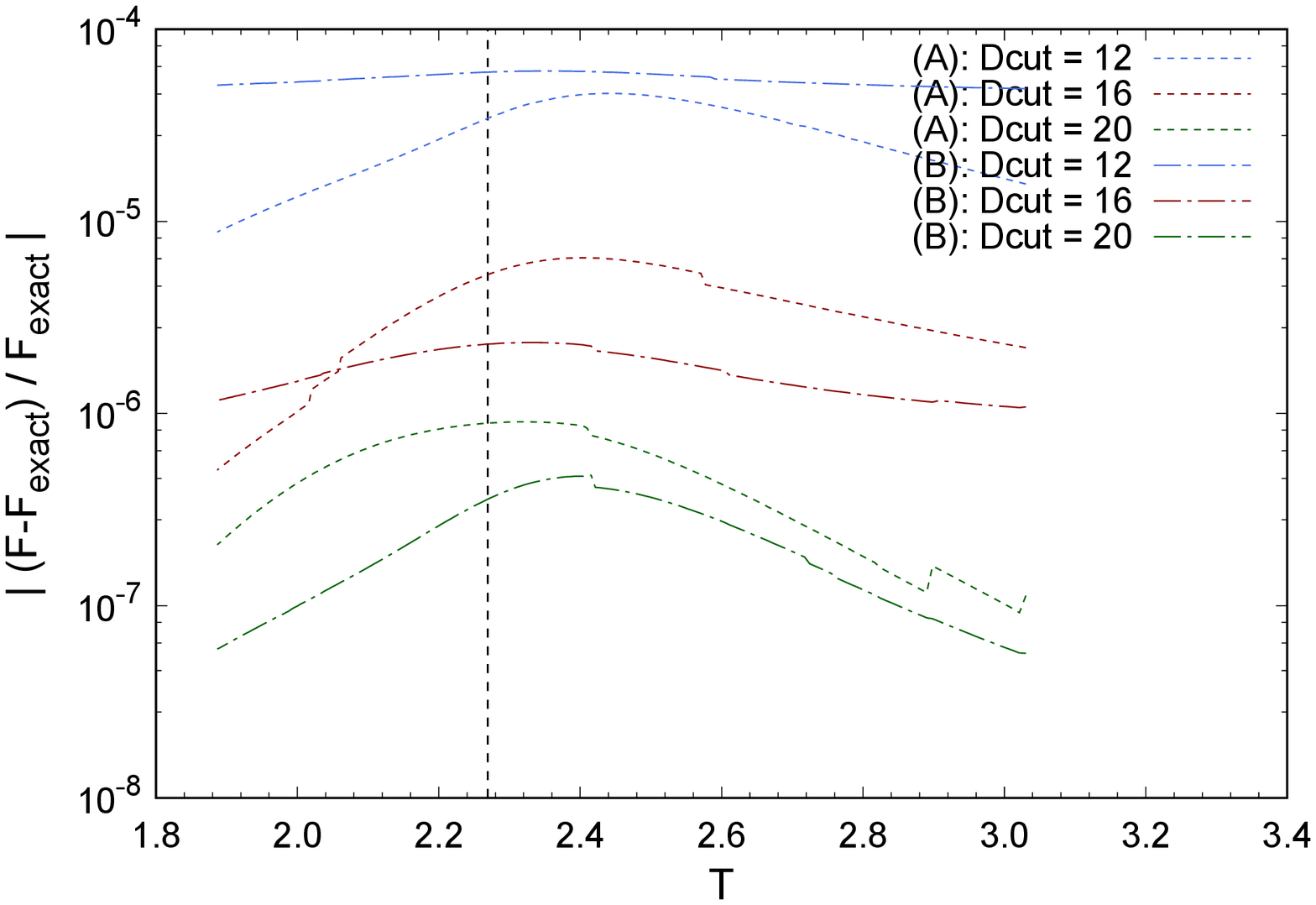}
		\caption{$T$ dependence of the relative residual of the free energy on $V=(16)^2$ lattice.
    The dotted line is for (A) the slanting-cut method with $\Delta=3$ 
   in Eq.~(\ref{eq:smoothcut_a}),
     and the chain line is for (B) the FDF-cut method with  $\sigma=1.0$ in Eq.~(\ref{eq:smoothcut_b}).}
		\label{smooth}
	\end{center}
\end{figure}

Figure~\ref{smooth} shows the relative residues obtained by these two cutoffs, which
show smoother temperature dependence compared to those in Fig.~\ref{crossover}.  
It is confirmed that the smooth cutoff scheme is effective to tame the irregular parameter dependence
found in the sharp cutoff scheme in the standard TRG method.

\section{Summary}
\label{sec:summary}

We have discussed  the issue of the irregular parameter dependence 
observed in the TRG results. 
We have investigated its origin using the two-dimensional Ising model and concluded that
the irregular behaviors is caused by the level crossing between the 
 singular values in the sharp cutoff scheme with $\Dcut$.
When the level crossing occurs between the $\Dcut$-th and $(\Dcut+1)$-th singular values, 
$\Dcut$-th singular vector is replaced by the completely different one across the crossover point, 
though the $\Dcut$-th singular value changes continuously as a function of the parameter. 
Thus the constructed tensor drastically changes and
yields the jump of the numerical result at the crossover point.

We have shown that the smooth cutoff  
improves the irregular behavior of the free energy in two-dimensional Ising model.
Further improvements would be important  to obtain the precise results in 
more complicated lattice models or higher dimensional models with the tensor network schemes.

\section*{Acknowledgments}
We would like to thank Ken-Ichi Ishikawa for encouraging our study.
This work was supported by
the Ministry of Education, Culture, Sports, Science and Technology (MEXT) as
 ``Exploratory Challenge on Post-K computer'' (Frontiers of Basic Science: Challenging the Limits),
and the Grant-in-Aid for JSPS Research Fellow (No.18J10663),
and JSPS KAKENHI Grant Numbers JP16K05328.


\bibliographystyle{unsrt}
\bibliography{refs}

\begin{thebibliography}{10}

\bibitem{Levin:2006jai}
Michael Levin and Cody~P. Nave.
\newblock {Tensor renormalization group approach to 2D classical lattice
  models}.
\newblock {\em Phys. Rev. Lett.}, 99(12):120601, 2007.

\bibitem{Shimizu:2012wfa}
Yuya Shimizu.
\newblock {Analysis of the $(1+1)$-dimensional lattice $\phi^{4}$ model using
  the tensor renormalization group}.
\newblock {\em Chin. J. Phys.}, 50:749, 2012.

\bibitem{Shimizu:2014uva}
Yuya Shimizu and Yoshinobu Kuramashi.
\newblock {Grassmann tensor renormalization group approach to one-flavor
  lattice Schwinger model}.
\newblock {\em Phys. Rev.}, D90(1):014508, 2014.

\bibitem{Unmuth-Yockey:2014iga}
Judah Unmuth-Yockey, Yannick Meurice, James Osborn, and Haiyuan Zou.
\newblock {Tensor renormalization group study of the 2d O(3) model}.
\newblock 2014.

\bibitem{Shimizu:2014fsa}
Yuya Shimizu and Yoshinobu Kuramashi.
\newblock {Critical behavior of the lattice Schwinger model with a topological
  term at $\theta=\pi$ using the Grassmann tensor renormalization group}.
\newblock {\em Phys. Rev.}, D90(7):074503, 2014.

\bibitem{Takeda:2014vwa}
Shinji Takeda and Yusuke Yoshimura.
\newblock {Grassmann tensor renormalization group for the one-flavor lattice
  Gross-Neveu model with finite chemical potential}.
\newblock {\em PTEP}, 2015(4):043B01, 2015.

\bibitem{Kawauchi:2016xng}
Hikaru Kawauchi and Shinji Takeda.
\newblock {Tensor renormalization group analysis of CP(N-1) model}.
\newblock {\em Phys. Rev.}, D93(11):114503, 2016.

\bibitem{Meurice:2016mkb}
Y.~Meurice, A.~Bazavov, Shan-Wen Tsai, J.~Unmuth-Yockey, Li-Ping Yang, and Jin
  Zhang.
\newblock {Tensor RG calculations and quantum simulations near criticality}.
\newblock {\em PoS}, LATTICE2016:325, 2016.

\bibitem{Sakai:2017jwp}
Ryo Sakai, Shinji Takeda, and Yusuke Yoshimura.
\newblock {Higher order tensor renormalization group for relativistic fermion
  systems}.
\newblock {\em PTEP}, 2017(6):063B07, 2017.

\bibitem{Yoshimura:2017jpk}
Yusuke Yoshimura, Yoshinobu Kuramashi, Yoshifumi Nakamura, Shinji Takeda, and
  Ryo Sakai.
\newblock {Calculation of fermionic Green functions with Grassmann higher-order
  tensor renormalization group}.
\newblock {\em Phys. Rev.}, D97(5):054511, 2018.

\bibitem{Shimizu:2017onf}
Yuya Shimizu and Yoshinobu Kuramashi.
\newblock {Berezinskii-Kosterlitz-Thouless transition in lattice Schwinger
  model with one flavor of Wilson fermion}.
\newblock {\em Phys. Rev.}, D97(3):034502, 2018.

\bibitem{Kuramashi:2018mmi}
Yoshinobu Kuramashi and Yusuke Yoshimura.
\newblock {Three-dimensional finite temperature Z$_2$ gauge theory with tensor
  network scheme}.
\newblock 2018.

\bibitem{Kadoh:2018hqq}
Daisuke Kadoh, Yoshinobu Kuramashi, Yoshifumi Nakamura, Ryo Sakai, Shinji
  Takeda, and Yusuke Yoshimura.
\newblock {Tensor network formulation for two-dimensional lattice $ \mathcal{N}
  $ = 1 Wess-Zumino model}.
\newblock {\em JHEP}, 03:141, 2018.

\bibitem{PhysRev.65.117}
Lars Onsager.
\newblock Crystal statistics. i. a two-dimensional model with an order-disorder
  transition.
\newblock {\em Phys. Rev.}, 65:117--149, Feb 1944.

\bibitem{PhysRev.76.1232}
Bruria Kaufman.
\newblock Crystal statistics. ii. partition function evaluated by spinor
  analysis.
\newblock {\em Phys. Rev.}, 76:1232--1243, Oct 1949.

\end{thebibliography}

\end{document}